\documentclass[%
 reprint,
superscriptaddress,
preprintnumbers,
nofootinbib,
 amsmath,amssymb,
]{revtex4-2}

\usepackage{dblfloatfix}
\usepackage{caption}
\usepackage{xcolor}
\usepackage{booktabs} 
\usepackage{graphicx}
\usepackage{hyperref}

\newcommand{\wmap}[0]{\textit{WMAP}}
\newcommand{\planck}[0]{\textit{Planck}}

\newcommand{\npipe}[0]{\texttt{NPIPE}}

\begin{document}

\title{Cosmic birefringence from a joint analysis of ACT and \planck}

\author{Johannes R. Eskilt}
\email{j.r.eskilt@astro.uio.no}
\affiliation{Institute of Theoretical Astrophysics, University of Oslo, P.O. Box 1029 Blindern, N-0315 Oslo, Norway}

\date{\today}
\begin{abstract}
Dark matter and dark energy could be ultra-light pseudoscalar fields that couple to electromagnetism through a Chern-Simons term. This would cause a parity-breaking rotation of the plane of linearly polarized light from the cosmic microwave background, known as cosmic birefringence, which is not predicted by $\Lambda$CDM. Recent hints of a non-zero cosmic birefringence angle have been found in both the \planck\, Data Release 4 and ACT Data Release 6 datasets independently. This work jointly analyzes the polarized maps of both, leveraging the cross-power spectra between the two telescopes. Taking the reported instrumental priors at face value, the joint analysis yields $\beta = 0.277^{\circ} \pm 0.057^{\circ}$, which is non-zero with a statistical significance of $4.8\sigma$. A robustness test mitigating contamination from dust emission excludes $\beta = 0^{\circ}$ at $3.5\sigma$, and retrieving $\beta=0^{\circ}$ would require both the dust $EB$ modeling and instrumental priors to fail simultaneously. However, unresolved systematics in the data must be understood before we can draw strong cosmological conclusions.
\end{abstract}

\maketitle
\section{\label{sec:intro}Introduction}
The $\Lambda$CDM theory represents our best understanding of the Universe, explaining most of our observations to a remarkable accuracy. But as it is a parity-symmetric theory, it allows us to search for parity-violating extensions that could shed light on the dark sector \cite{komatsu:2022}. One such extension is a Chern-Simons coupling between electromagnetism and an axion-like field, causing the linear polarization of light to rotate as it travels through the Universe \cite{turner/widrow:1988, carroll/field/jackiw:1990, harari/sikivie:1992, carroll:1998, fujita/etal:2021b}.

We probe this cosmic birefringence angle $\beta$ through the correlation between $E$-modes and $B$-modes of the linearly polarized light from the cosmic microwave background (CMB) \cite{seljak/zaldarriaga:1997, kamionkowski/etal:1997}. A non-zero cosmic birefringence signal would mix these intrinsic $E$-modes and $B$-modes \cite{lue/wang/kamionkowski:1999}, with the observed spherical harmonic coefficients of the CMB to become $E^{\rm o}_{\ell m} = E^{\rm CMB}_{\ell m}\cos(2\beta) + B^{\rm CMB}_{\ell m}\sin(2\beta)$ and $B^{\rm o}_{\ell m} = E^{\rm CMB}_{\ell m}\sin(2\beta) + B^{\rm CMB}_{\ell m}\cos(2\beta)$. Combining this into the observed $EB$ power spectrum gives us:
\begin{equation}
    \label{eq:simple_birefringence}
    C^{EB,\,\rm o}_\ell = \frac{\tan(4\beta)}{2}\left(C^{EE,\, \rm o}_{\ell} - C^{BB,\, \rm o}_{\ell} \right) + \frac{C^{EB,\, \rm CMB}_\ell}{\cos(4\beta)},
\end{equation}
where $C^{EB,\, \rm CMB}_\ell$ is the intrinsic $EB$ correlation of the CMB which is not predicted by $\Lambda$CDM. We therefore set it to zero in this work \cite{Fujita:2022qlk}. The $E$-modes of the CMB dominate over $B$-modes, meaning cosmic birefringence can be measured by comparing the shape of $C^{EB,\, \rm o}_\ell$ to $C^{EE,\,\rm o}_\ell$.

Systematics such as the miscalibration angle $\alpha$ complicate this measurement. Imperfect calibration causes the instrument to measure the wrong polarization angle. This global rotation angle leads us to replace $\beta \rightarrow \alpha + \beta$ in Eq.~\eqref{eq:simple_birefringence}. In practice, without a constraint on $\alpha$, you cannot break the $\alpha + \beta$ degeneracy.

\begin{figure}
\centering
\includegraphics[width=\linewidth]{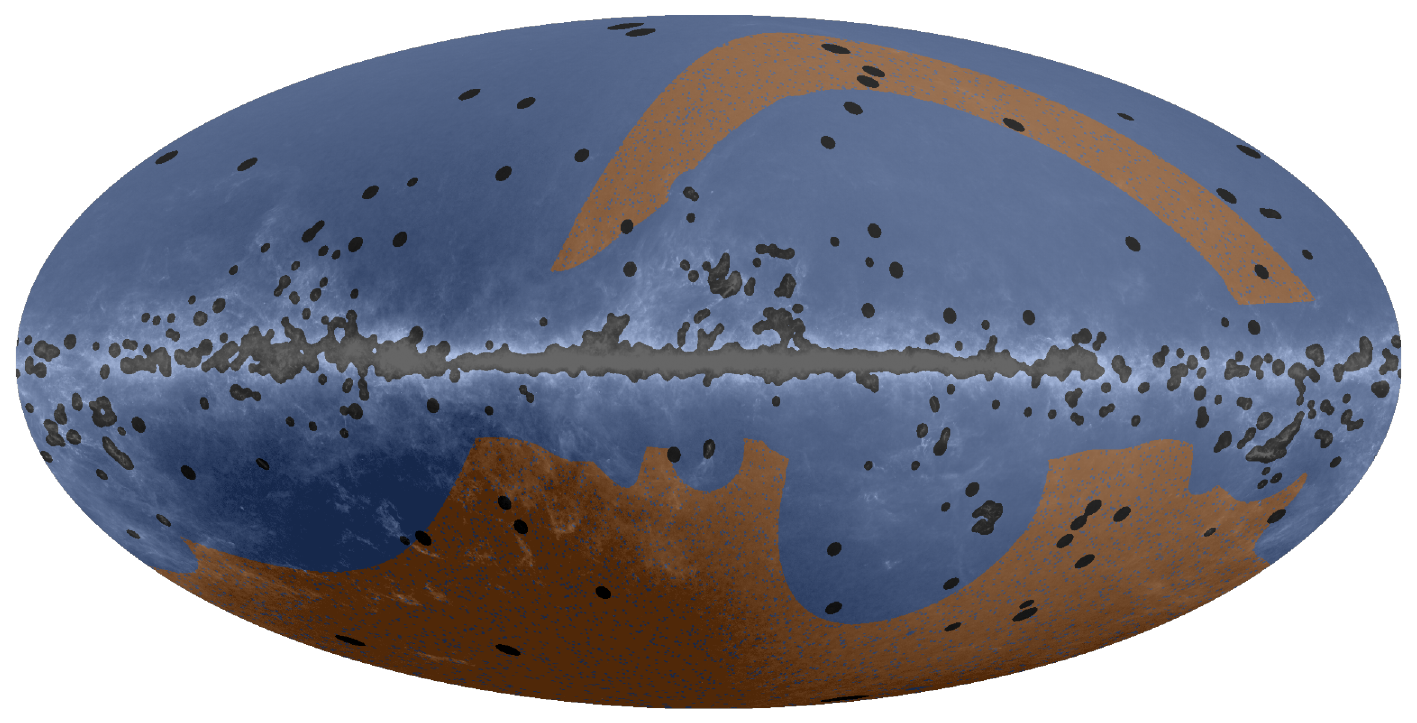}
\captionsetup{justification=raggedright}
\caption{\label{fig:masks} The observation window of ACT in orange with the nearly full-sky mask used for \planck\, in gray. The background shows the temperature map of the 353\,GHz band for illustrative purposes.
}
\end{figure}

Precise on-ground calibration of the instrument or calibrating using the polarized foreground emission are two ways of determining $\alpha$. The latter method was first developed in Ref.~\cite{minami/etal:2019} and further generalized in Refs.~\cite{minami:2020, minami/komatsu:2020}. The authors realized that if cosmic birefringence rotates the plane of the linearly polarized light from the CMB on a cosmological time scale, then the polarized foreground emission from the Milky Way will be negligibly impacted by $\beta$. In other words, the CMB photons are impacted by $\alpha + \beta$ but the Galactic foreground emission is only rotated by $\alpha$.

Decomposing the observed $E$ and $B$ modes into foreground and CMB components, we arrive at \cite{minami/etal:2019}
\begin{align}
    \label{eq:complicated_birefringence}
    \nonumber
    C^{EB,\,\rm o}_\ell &= \frac{\tan(4\alpha)}{2}\left(C^{EE,\, \rm o}_{\ell} - C^{BB,\, \rm o}_{\ell} \right) + \frac{C^{EB, \, \rm fg}_\ell}{\cos(4\alpha)}\\
    &+\frac{\sin(4\beta)}{2\cos(4\alpha)}\left(C^{EE,\, \rm CMB}_{\ell} - C^{BB,\, \rm CMB}_{\ell} \right),
\end{align}
where $C^{EB, \, \rm fg}_\ell$ is the intrinsic $EB$ power spectrum of the Galactic foreground emission.

The polarized \planck\,data has been extensively studied for cosmic birefringence since the work of Ref.~\cite{minami/etal:2019}. First for the High Frequency Instrument (HFI) of \planck\, \cite{minami/komatsu:2020b, NPIPE:2022} and then combined with the Low Frequency Instrument (LFI) \cite{Eskilt:2022wav, Remazeilles:2025wzd} and \wmap~\cite{Eskilt:2022cff, Cosmoglobe:2023pgf}, with the strongest measurement coming from the joint analysis of \planck\, and \wmap\, with $\beta = 0.342_{\,\,\,-0.091^{\circ}}^{\circ +0.094^{\circ}}$, excluding zero at $3.6\sigma$ \cite{Eskilt:2022cff}. We report uncertainties with 68\% confidence level in this work. This measurement used a filamentary model for dust $EB$ correlations inspired by Ref.~\cite{clark/etal:2021} which increased the significance of $\beta$.

In March 2025, the ACT team published their Data Release 6 results \cite{AtacamaCosmologyTelescope:2025blo, AtacamaCosmologyTelescope:2025vnj}. Instead of directly measuring cosmic birefringence, they estimated the rotation angle $\psi_i = \alpha_i + \beta$ for each band $i$ and averaged them to get $\langle \psi\rangle = 0.20^{\circ} \pm 0.08^{\circ}$, an increase from previous ACT measurements \cite{namikawa/etal:2020, choi/etal:2020}. The error bar included the systematic uncertainty of $\alpha_i$ derived from the optics model of the instruments.

Ref.~\cite{Diego-Palazuelos:2025dmh} instead used a Bayesian framework to break the $\alpha+\beta$ degeneracy by setting Gaussian priors on $\alpha_i$ from the calibration uncertainties \cite{Murphy:2024fna}. Additionally, they used $TB$ data and PA4 220\,GHz band which was discarded by the ACT team due to its low constraining power. They obtained a comparable $\beta = 0.215^\circ \pm 0.074^\circ$. 

The \planck\, and ACT datasets have so far only been analyzed independently for isotropic cosmic birefringence. We perform a joint analysis because the two break the $\alpha+\beta$ degeneracy in independent ways: the former with the polarized foreground emission and the latter with priors on $\alpha$ alone from an optics model as the ACT analysis masks the Galactic plane and probes only $\ell > 600$. The dominant systematic vulnerability of each dataset is not an input the other one relies on. We can additionally probe the correlations at the overlapping multipoles, $600 \lesssim \ell \lesssim 2000$, where the noise and instrumental systematics are uncorrelated.

\section{\label{sec:pipeline}Data and Method}

\planck\,was an ESA space telescope operating between 2009 and 2013 \cite{Planck2018I}, carrying two instruments developed and calibrated by different teams: the High Frequency Instrument \cite{Planck2018III} which could measure the polarized sky at 100, 143, 217, and 353\,GHz, and the Low Frequency Instrument \cite{Planck2018II} with the 30, 44, and 70\,GHz bands.

We use the publicly available maps of the \planck\, Data Release 4, also known as the \npipe\, release~\cite{NPIPE:2022}. They split the detectors per frequency into two groups, creating two A/B detector split maps per band, giving one $\alpha_i$ per data split $i$. 30 and 44\,GHz bands have one $\alpha_i$ each as these use half-mission maps instead. The data products are publicly available\footnote{\url{https://pla.esac.esa.int/pla/}}.

The Atacama Cosmology Telescope Data Release 6 (ACT DR6) measured 19,000 square degrees of the sky between 2017 and 2022 with three optics tubes, each housing a polarization array (PA) \cite{AtacamaCosmologyTelescope:2025blo}. The mid-frequency arrays, PA5 and PA6, contained two bands each, 77-112\,GHz (f090) and 124-172\,GHz (f150), while the high-frequency array PA4 had an f150 and a 182-227\,GHz (f220) band. The PA4 f150 band failed null tests and was not used for cosmological analysis by ACT, so we will discard it but use the other bands. The team released four time-split maps per band that were developed to have similar noise characteristics \cite{AtacamaCosmologyTelescope:2025vnj}. As they are time-splits, each ACT band has its own miscalibration angle. The ACT DR6 release included both data products\footnote{\url{https://lambda.gsfc.nasa.gov/product/act/act_dr6.02/}} and publicly available code to reproduce their results\footnote{\url{https://github.com/simonsobs/PSpipe}}$^{,}$\footnote{\url{https://github.com/simonsobs/pspipe_utils}}.

We will follow the cosmic birefringence methodology of Ref. \cite{Eskilt:2022cff, minami/komatsu:2020} which includes a dust $EB$ model. The likelihood is
\begin{equation}
    \label{eq:likelihood}
    \ln L = -\frac{1}{2} \sum_{b} \left(\vec{v}_b^T \textbf{M}_b^{-1}\vec{v}_b + \ln |\textbf{M}_b| +\vec{\alpha}^{T} \Pi^{-1}\vec{\alpha}\right)\,,
\end{equation}
where $b$ is the center of a multipole bin, $\textbf{M}_b = \textbf{A}_b\text{Cov}(\vec{C}^{\textrm{o}}_b, \vec{C}^{\textrm{o}}_b)\textbf{A}_b^T$ and  $\vec{v}^T_b \equiv \textbf{A}_b\vec{C}^{\textrm{o}}_b - \textbf{B}_b\vec{C}^{\Lambda \text{CDM}}_b$. $\vec{v}^T_b = 0$ is a multi-band generalization of Eq.~\eqref{eq:complicated_birefringence}. The observed spectra are $\vec{C}^{\textrm{o}}_b = \begin{bmatrix}
  C^{E_i E_j,\, \textrm{o}}_b, C^{B_i B_j, \, \textrm{o}}_b, C^{E_i B_j,\, \textrm{o}}_b
\end{bmatrix}^T$ and the $\Lambda$CDM spectra are $\vec{C}^{\Lambda \text{CDM}}_b = \left [ C^{E_iE_j,\,\Lambda \text{CDM}}_{b}, C^{B_iB_j,\,\Lambda \text{CDM}}_{b} \right]$. We refer the reader to Ref.~\cite{Eskilt:2022cff} for the definitions of the matrices $\textbf{A}_b$ and $\textbf{B}_b$. Both matrices incorporate a filamentary dust model used in Refs.~\cite{NPIPE:2022, Eskilt:2022wav, Eskilt:2022cff} inspired by the work of Refs.~\cite{clark/etal:2021, huffenberger/rotti/collins:2020},
\begin{equation}
    \label{eq:dust_ansatz}
     C^{EB, \, \text{dust}}_\ell = A_\ell C^{EE,\, \text{dust}}_\ell \sin \left(4\psi_\ell\right)\,.
\end{equation}

Here, $\psi_\ell = \frac12\arctan\left(C^{TB,\,\rm dust}_\ell / C^{TE,\,\rm dust}_\ell\right)$ is a multipole-averaged misalignment angle between the dust filaments and Galactic magnetic fields, and the parameter $A_\ell$ is expected to be weakly varying.

We estimate $\psi_\ell$ from the dust-dominated 353\,GHz $TB$ and $TE$ spectra following previous work \cite{Eskilt:2022wav, NPIPE:2022}, using the split orderings recommended by Ref.~\cite{Hervias-Caimapo:2024ili} to mitigate instrumental systematics: $T_{353B} \times E_{353A}$ for TE and $T_{353A} \times B_{353B}$ for TB, smoothed with a Gaussian filter. As Ref.~\cite{Hervias-Caimapo:2024ili} found a growing bias in $TB$ at $\ell > 600$ and foreground is negligible to the CMB at high multipoles, we apply the dust model to HFI-only spectra at bin centers $b < 600.5$. We sample three $A_\ell$ amplitudes in the ranges $56 \leq \ell \leq 135$, $136 \leq \ell \leq 215$, $216 \leq \ell \leq 575$ with flat positive priors. The LFI bands are foreground-dominated by synchrotron emission, and we assume $EB$ correlation of synchrotron is zero as there is no evidence for it \cite{Martire:2021gbc, Rubino-Martin:2023fya}.

We use \texttt{CAMB}\footnote{\url{https://github.com/cmbant/CAMB}}~\cite{Lewis:2000} with the \planck\,2018 best-fit cosmological parameters\,\citep{Planck2018VI} to generate the $\Lambda$CDM spectra. We do not beam-deconvolve the observed spectra, so we multiply the \texttt{CAMB} spectra by the respective beam transfer functions and pixel window functions.

We bin $\Delta\ell = 20$ for $56\leq\ell < 576$, $\Delta\ell = 50$ for $576 \leq \ell \leq 2025$, increasing to $\Delta\ell = 800$ at $\ell > 6325$. For $\ell\geq576$, this follows the same binning as ACT DR6, while keeping $\Delta \ell = 20$ at lower multipoles to follow a similar binning used in earlier \planck\, analyses \cite{minami/komatsu:2020b, Eskilt:2022cff, Eskilt:2022wav, NPIPE:2022}. For the ACT bands, we follow the baseline multipole ranges set by the ACT team based on null tests \cite{AtacamaCosmologyTelescope:2025vnj}, and additionally set $b_{\rm min} = 1000.5$ for PA4 f220\,\cite{Diego-Palazuelos:2025dmh}. We use $65.5 \leq b \leq 2000.5$ for \planck.

For \planck, we discard split auto spectra, like 100A $\times$ 100A, to avoid noise contamination. For ACT, we instead follow Appendix C.4 in Ref.~\cite{AtacamaCosmologyTelescope:2025vnj} and average over a combination of four time-split maps, excluding time-split auto-correlations, to arrive at band auto-power spectra, like PA5 f090 $\times$ PA5 f090.

The size of the vector $\vec{v}^T_b$ and covariance matrix $\textbf{M}_b$ in Eq.~\eqref{eq:likelihood} depends on the number of valid bands for the given bin $b$.   For example, we utilize all bands for bin center $b = 1000.5$, meaning we use the $EB$ power spectra of 25 ACT $\times$ ACT ($5^2$ spectra), 140 ACT $\times$ \planck\,(2 $EB$ permutations $\times$ 5 ACT bands $\times$ 14\,\planck\, data splits), and 182 \planck\, $\times$ \planck\, ($14^2$ spectra  - 14 auto-correlations), a total of 347 different $C^{EB,\,\rm o}_\ell$.

We use the same nearly full-sky mask for \planck\, as in Ref.~\cite{Eskilt:2022wav} which has excluded the union of the point sources for all the polarized \planck\, bands. It also excludes carbon-monoxide (CO) line stronger than $45\,\textrm{K}_{\textrm{RJ}}\, \textrm{km} \,\textrm{s}^{-1}$ as this can cause temperature-to-polarization leakage. The resulting mask has a sky fraction of $f_{\rm sky} = 0.92$. For ACT, we use the masks released by the ACT team which removes point sources and the Galactic center, giving $f_{\rm sky} = 0.25$. For ACT $\times$ \planck, we use a combination of the point source mask of \planck\, with the observation window of PA5 f090. These masks can be seen in Fig.~\ref{fig:masks}.

The only addition to Eq.~\eqref{eq:likelihood} compared to Ref.~\cite{Eskilt:2022cff} is the Gaussian prior $\Pi$ set on the miscalibration vector $\vec{\alpha}^T = (\alpha_i)$. We set the same priors on ACT as Ref.~\cite{Diego-Palazuelos:2025dmh}, namely we set the mean of $\alpha_i$ to zero, $\sigma_{\alpha_{\rm PA4}} = 0.09^{\circ}$, $\sigma_{\alpha_{\rm PA5}} = 0.09^{\circ}$, and $\sigma_{\alpha_{\rm PA6}} = 0.11^{\circ}$, which was determined in Ref.~\cite{Murphy:2024fna} using an optics model of the telescope and instruments. In addition, we follow the baseline result of Ref.~\cite{Diego-Palazuelos:2025dmh} which assumes a $90\%$ correlation between miscalibration angles of bands in the same polarization array as they share the same optics tube.

While we rely on the foreground emission to constrain $\alpha$ for \planck, we also add \planck\, priors on $\alpha_i$. This will have less constraining power than the foreground emission but allows us to constrain $\alpha$ when we later mask the Galactic plane. Ref.~\cite{Rosset2010} reported pre-flight an upper limit of $0.3^{\circ}$ on the absolute orientation uncertainty of the focal plane of the HFI, and a relative uncertainty between individual polarization-sensitive bolometers (PSBs) to have an upper limit of $0.9^{\circ}$. We turn this into priors on $\alpha_i$ by following the \planck\, Collaboration \cite{PlanckIntXLIX}. Assuming these bounds are uniform, we find the standard deviations to be $\sigma_{\rm rel} = 0.9^{\circ}/\sqrt{3} \approx 0.52^{\circ}$ and $\sigma_{\rm abs} = 0.17^{\circ}$. Noting that the relative uncertainty can be averaged over the 4 PSBs that make up each HFI data split map, the priors on HFI $\alpha_i$ become $\sigma_{\alpha_i} = \sqrt{\sigma_{\rm abs}^2 + \sigma_{\rm rel}^2/4} = 0.31^{\circ}$.

\begin{figure}
\centering
\includegraphics[width=\linewidth]{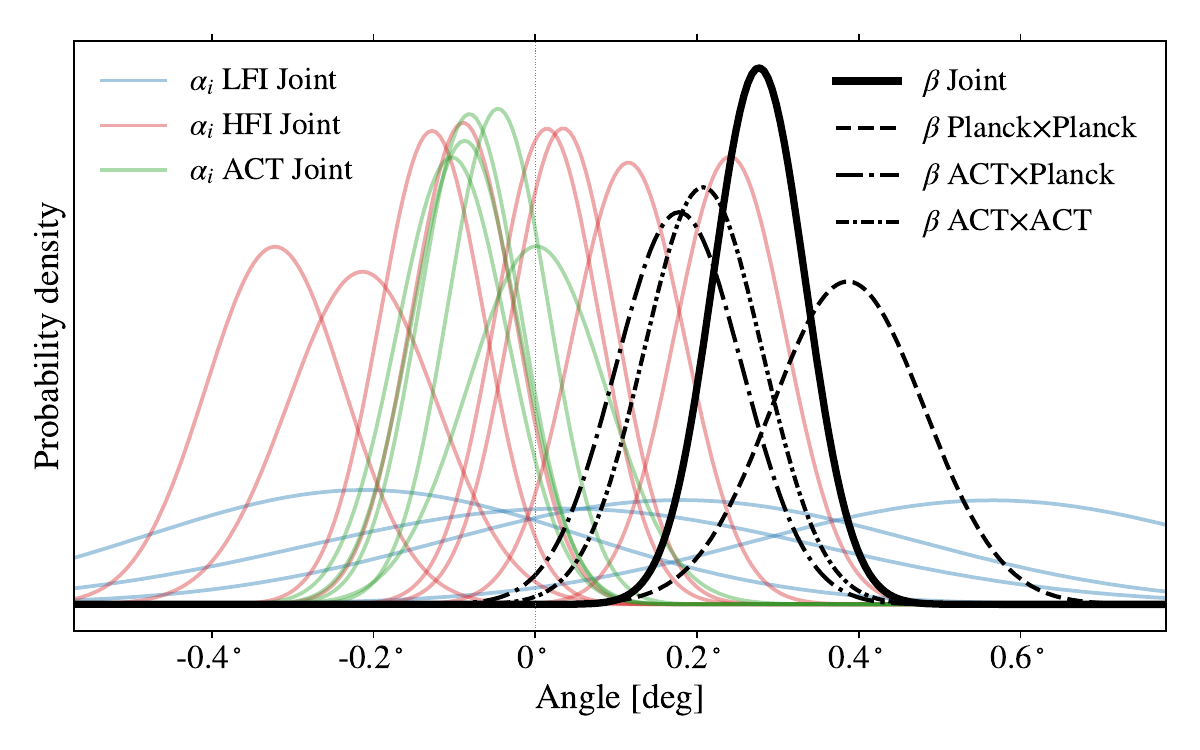}
\captionsetup{justification=raggedright}
\caption{\label{fig:posteriors} Posterior distribution of the miscalibration angles $\alpha_i$ and the cosmic birefringence angle $\beta$ from the joint analysis of ACT and \planck\, in solid lines. We also show measurements of $\beta$ from data subsets in non-solid lines.
}
\end{figure}
As all the HFI detectors share the same uncertainty in the absolute orientation of the focal plane, the correlation between the miscalibration angle for each data split map becomes $\rho = \sigma_{\rm abs}^2/{\sigma_{\alpha_i}^2} = 0.31$. We approximate the $\sigma_{\alpha_i}$ and $\rho$ as a multivariate Gaussian distribution of $\alpha_i$ with mean of zero and use it as our prior in Eq.~\eqref{eq:likelihood}. The polarization angles of LFI were poorly constrained pre-flight, and we put a conservative $1^{\circ}$ Gaussian prior and assume no correlation between the bands \cite{2010A&A...520A...8L}.

We develop a pipeline that calculates the power spectrum and covariance from the released maps. The \planck-only spectra and covariances follow previous work \cite{minami/komatsu:2020b, NPIPE:2022, Eskilt:2022wav, Eskilt:2022cff}, but the ACT maps need more care as they have more complicated noise properties and smaller sky coverages. The details of the computation of the spectra and covariance can be found in Appendix A.

We find good agreement between the released spectra and ours but the $EB$ uncertainty is underestimated by $7\%$ compared to the covariance of ACT (see Fig.~\ref{fig:spectra}).

Our baseline result will use the released ACT DR6 spectra and covariance for the ACT $\times$ ACT block, but for any cross-power spectra or covariance elements with at least one \planck\,map, we will rely on our own pipeline. The consistency between the official ACT pipeline and ours gives confidence in the cross-experiment products. Our pipeline code is publicly available\footnote{\href{https://github.com/LilleJohs/cosmic-birefringence-planck-act}{github.com/LilleJohs/cosmic-birefringence-planck-act}}.

\section{\label{sec:results}Results}
We first verify our cosmic birefringence pipeline by comparing our ACT $\times$ ACT results with Ref.~\cite{Diego-Palazuelos:2025dmh}. Using the published ACT spectra and covariance, we find $\beta = 0.207^{\circ} \pm 0.073^{\circ}$, in great agreement with Ref.~\cite{Diego-Palazuelos:2025dmh}. Unlike their analysis, we exclude $TB$ correlations and off-diagonal bin-to-bin covariance elements. Both choices have negligible impact: Ref.~\cite{Diego-Palazuelos:2025dmh} cites only a 1.24\% tightening of constraints when $TB$ is included, and we find that neglecting the off-diagonal bin-to-bin elements increases the $\beta$ uncertainty by only ${\approx}0.001^{\circ}$ and bias the parameters $<0.1\sigma$. We also do not correct for residual temperature-to-polarization leakage, as the authors concluded that it has no significant impact on their results.

The $7\%$ underestimation of the $EB$ uncertainty in ACT $\times$ ACT does not affect our baseline, which uses the released DR6 covariance for the ACT-block, but it may affect cross-elements with \planck. Estimating $\alpha_i$ and $\beta$ on the ACT-only products from our pipeline gives $\beta = 0.205^{\circ}\pm 0.072^{\circ}$, an underestimation of $2\%$ of $\sigma_{\beta}$ with $<0.1\sigma$ parameter bias, since the $\alpha_i$ priors partially absorb the missing $EB$ variance. We conclude the effect on $\beta$ is insignificant.     

Turning to our \planck-only pipeline, we find excellent agreement with Ref.~\cite{Eskilt:2022cff, NPIPE:2022}. Unlike previous work that analyzed \planck, we use a wider bin of $\Delta \ell = 50$ at $\ell > 600$ causing slightly larger uncertainties in the estimated parameters, but the new priors we place on $\alpha_i$ give a total tightening of $\sigma_{\beta}$ by $6\%$ compared to Ref.~\cite{Eskilt:2022wav}. The nearly full-sky measurement with \planck-only spectra gives $\beta = 0.387^{\circ} \pm 0.094^{\circ}$ using the filamentary model for dust $EB$. This is in $1.5\sigma$ tension with the ACT-only results, which could be due to \planck\,having a stronger $EB$ signal giving $\alpha+\beta \approx 0.30^\circ$ \cite{PlanckIntXLIX, NPIPE:2022} while ACT prefers $\alpha+\beta \approx 0.20^{\circ}$ \cite{AtacamaCosmologyTelescope:2025blo, Diego-Palazuelos:2025dmh}. The two measurements also use different methods to break $\alpha+\beta$ degeneracy: \planck\, uses the foreground emission, while ACT relies on $\alpha_i$ priors. An imperfect dust $EB$ model or priors could bias either leg.

We then turn to the ACT $\times$ \planck\,cross-power spectra, for example $E_{\rm PA6\,f150}B_{100A}$ and $E_{100A}B_{\rm PA6\,f150}$, which have overlapping multipoles in the range $600 < \ell < 2000$, and we find $\beta = 0.178^{\circ}\pm0.078^{\circ}$ with a significance of $2.3\sigma$, aligning closely with the ACT-only results. As there is little foreground emission both due to the Galactic cut mask and the higher multipole range, the $\alpha+\beta$ degeneracy is mainly broken by the ACT priors on $\alpha_i$, which are $\approx3$ times tighter than the HFI priors.

\begin{figure}
\centering
\includegraphics[width=\linewidth]{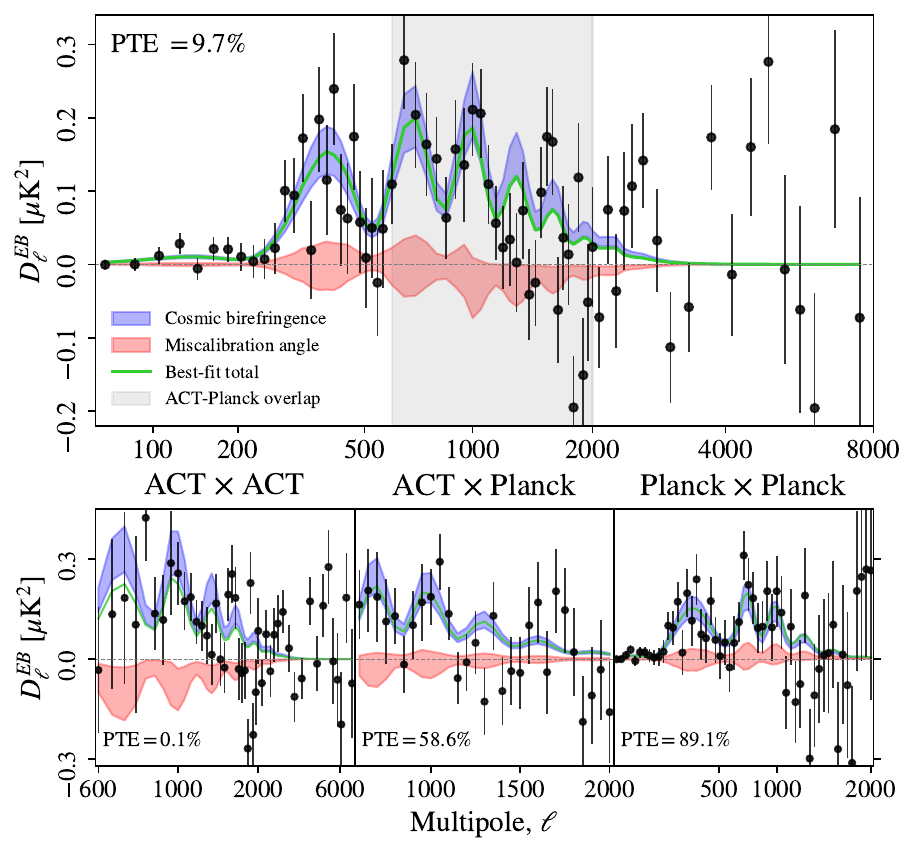}
\captionsetup{justification=raggedright}
\caption{\label{fig:stacked_eb} Inverse-variance weighted average (stacked) of the beam-convolved $EB$ power spectra for the joint analysis of ACT and \planck\,in the top panel. The gray area shows the multipole overlap between the two experiments. The bottom panels show the stacked $EB$ for the three data subsets. 
}
\end{figure}
We then perform our baseline result which is the full joint analysis, including ACT $\times$ ACT, ACT $\times$ \planck, \planck\,$\times$\,\planck\,spectra, and their respective covariances. We find $\beta = 0.277^{\circ} \pm 0.057^{\circ}$ which excludes $\beta=0$ at $4.8\sigma$. We show the estimated miscalibration angles and the cosmic birefringence angle in Fig.~\ref{fig:posteriors}. We also show the estimated $\beta$ from the three data subsets in non-solid lines, demonstrating the consistency of the two datasets preferring a positive $\beta$.

The ACT power spectra are dominated by CMB and noise over foreground as they mask the Galaxy and are only sensitive to $\ell > 600$. The \planck-only spectra in our baseline results keep most of the Galactic center and use the foreground polarization at lower $\ell$ to break the $\alpha+\beta$ degeneracy. As a robustness test to mitigate the effect of unknown $C^{EB,\,\rm fg}_\ell$, we follow the same principle on \planck\, data: we compute the \planck\, spectra on the $30\%$ Galactic cut mask used in Ref.~\cite{Eskilt:2022wav} with point sources and CO removed, yielding $f_{\rm sky} = 0.62$ and set $b_{\rm min} = 600.5$ for \planck. We additionally remove the 30, 44, and 353\, GHz bands as these are foreground-dominated. The ACT-leg remains unchanged, and sampling all miscalibration angles jointly with $\beta$, we find $\beta = 0.236^\circ \pm 0.067^\circ$ at $3.5\sigma$. This result is mostly determined by the tighter ACT priors on $\alpha_i$, but comparing this to the ACT-only results shows that the addition of \planck\, increases the significance of $\beta$. This measurement cannot be explained by foreground $EB$ correlations.

The $\alpha$ priors on HFI channels are derived from pre-flight calibrations \cite{Rosset2010} rather than in-flight and could be inaccurate. As robustness tests, we repeat the baseline analysis removing the \planck\, priors entirely, recovering $\beta = 0.280^{\circ} \pm 0.059^{\circ}$, and removing both the \planck\, and ACT $\alpha$ priors, which returns $\beta = 0.442^\circ \pm 0.098^\circ$, similar to the \planck-only analysis. In both cases, $\beta$ remains far from zero as the polarized foreground emission constrains $\alpha$ on the \planck-leg.

To visualize the positive $EB$ signal found in both experiments, we show the inverse-variance weighted $EB$ in Fig.~\ref{fig:stacked_eb} following Ref.~\cite{Eskilt:2022wav}. The top panel shows all data for the joint analysis, while the bottom panels decompose the signal into three blocks. The best-fit parameters give $\chi^2 = 89.1$ for the joint analysis, which has a probability to exceed (PTE) of $9.7\%$ with 73 degrees of freedom. This is reduced by the low $\rm{PTE} = 0.1\%$ in ACT $\times$ ACT for two reasons: it prefers a lower $\beta \approx 0.20^{\circ}$, and the ACT team found that $\langle \psi \rangle \approx 0.2^{\circ}$ reached PTE = $1\%$ \cite{AtacamaCosmologyTelescope:2025blo}, which they attributed to a poor fit around $\ell \approx 1800-1900$. For ACT $\times$ \planck\, and \planck\,$\times$ \planck, we find a good fit with the data giving $\textrm{PTE} = 58.6\%$ and $\textrm{PTE}=89.1\%$, respectively.

\section{\label{sec:conclusion}Conclusions}
This work performs a joint analysis of isotropic cosmic birefringence in the ACT DR6 and \planck\, DR4 datasets. Jointly sampling $17$ miscalibration angles, three dust $EB$ parameters, and one cosmic birefringence angle $\beta$, we find $\beta = 0.277^{\circ} \pm 0.057^{\circ}$, which excludes $\beta=0$ at $4.8\sigma$. This lies between the reported birefringence angles found in ACT and \planck\,separately \cite{Diego-Palazuelos:2025dmh, Eskilt:2022wav}.

Dividing our spectra into three blocks (ACT $\times$ ACT, ACT $\times$ \planck\, and \planck\, $\times$ \planck), we find positive $EB$ correlations not only in the joint data but in all three blocks individually as seen in Fig.~\ref{fig:stacked_eb}. We find constraining power in the ACT $\times$ \planck\, power spectra preferring $\beta = 0.178^{\circ}\pm0.078^{\circ}$ which has a significance of $2.3\sigma$. This aligns with the angle measured on the ACT $\times$ ACT block, $\beta = 0.207^{\circ} \pm 0.073^{\circ}$. However, it is in a mild tension with the $\beta = 0.387^{\circ} \pm 0.094^{\circ}$ preferred by \planck\,(see Fig.~\ref{fig:posteriors}).

\begin{figure}
\centering
\includegraphics[width=\linewidth]{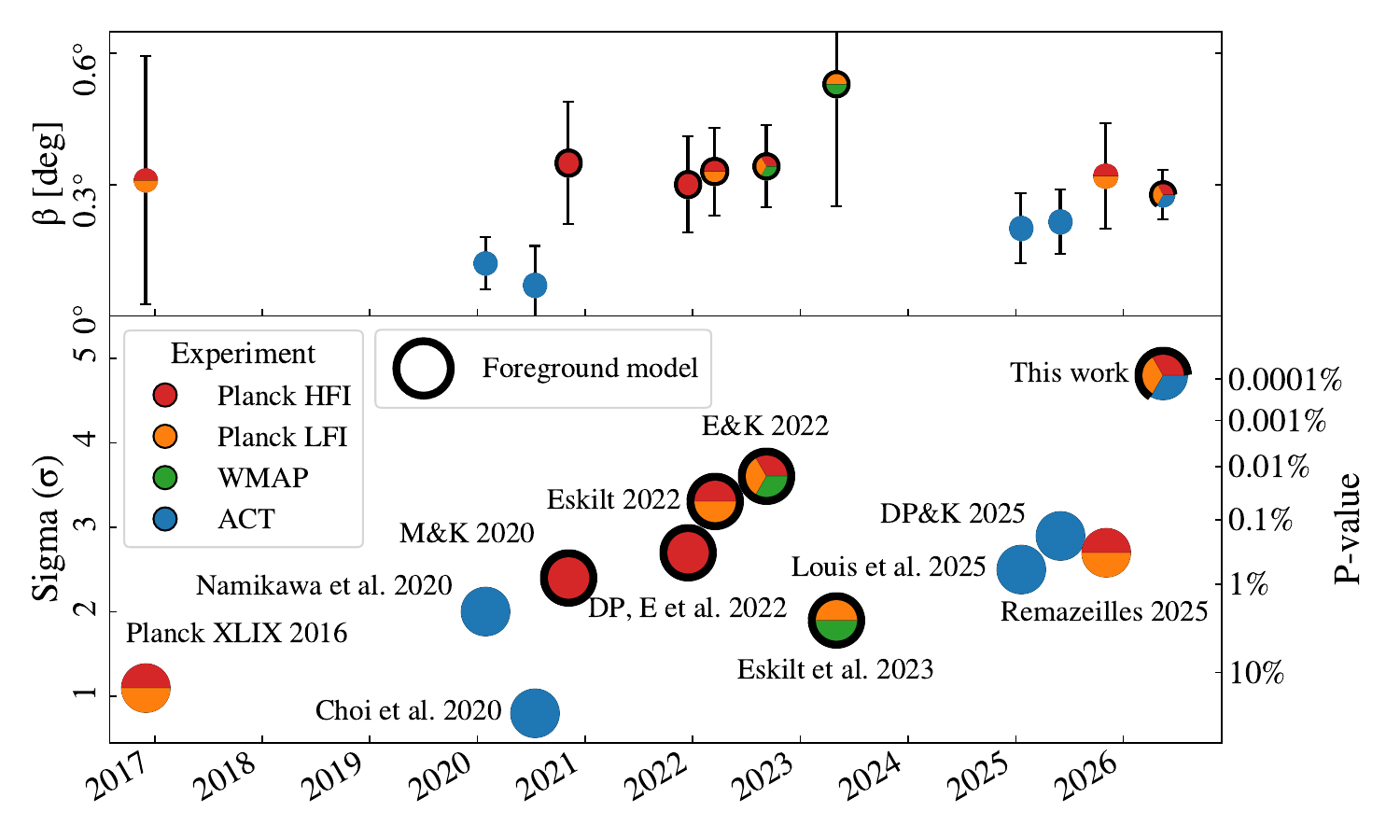}
\captionsetup{justification=raggedright}
\caption{\label{fig:birefringence_sigma} Timeline of cosmic birefringence measurements in \wmap, ACT\, and \planck. The upper panel shows $\beta$ measurements, and the lower shows the significances. The black rings indicate if a foreground $EB$ model was used.
}
\end{figure}

The $\alpha + \beta$ degeneracy is broken independently in the two datasets with instrumental priors on the ACT-leg and polarized foreground emissions for \planck. Although we model dust $EB$ correlations, incomplete knowledge of the foreground emissions means we cannot rule out a bias in $\beta$. To mitigate this, we mask \planck\, with a 30\% Galactic cut, probe only high multipoles $\ell > 600$ where CMB dominates over foreground emission, and we remove the foreground-dominated \planck\, spectra. We find $\beta = 0.236^\circ \pm 0.067^\circ$ which is non-zero with a statistical significance of $3.5\sigma$ and cannot be explained by foreground $EB$. This shows that $\beta = 0^\circ$ can only be recovered if both the $\alpha$ priors and the foreground $EB$ model fail.

Refs.~\cite{Cosmoglobe:2023pgf, NPIPE:2022} analyzed instrumental systematic effects on $\beta$ in \planck\, but could not explain the signal. On the ACT side, the team did not interpret their own measurement as cosmic birefringence, instead reporting an unexplained discrepancy in the polarization angle of PA5 f090 and PA5 f150~\cite{AtacamaCosmologyTelescope:2025blo, Murphy:2024fna}. Confirming the cosmic birefringence measurement will require a better understanding of both datasets, as unknown systematics in neither \planck, nor ACT can yet be excluded.

We summarize the past 10 years of isotropic cosmic birefringence measurements in ACT, \planck\, and \wmap\, in Fig.~\ref{fig:birefringence_sigma}. Here, we take each reported angle as a birefringence measurement, though the original analyses vary in how firmly they make that interpretation. In this period, the significance has increased, driven by improved data and the combination of datasets.

The signal has been shown to be consistent with frequency independence \cite{Eskilt:2022cff, Eskilt:2022wav}, the signature of an axion-like field coupled to electromagnetism via a Chern-Simons term. Confirmation of a cosmological origin will guide us toward a parity-violating theory of the Universe beyond $\Lambda$CDM, potentially shedding light on the dark sector.

\begin{acknowledgments}
We thank Sigurd Naess for help with understanding the ACT DR6 maps. We also thank Thibaut Louis for help with the ACT masks and Duncan Watts for feedback on an earlier draft. We are grateful for feedback on Fig.~\ref{fig:birefringence_sigma} from Eiichiro Komatsu. We acknowledge funding from the European Research Council (ERC) under the Horizon 2020 Research and Innovation Programme (Grant agreement No. 819478). \planck\, is a project of the European Space Agency (ESA) with instruments provided by
two scientific consortia funded by ESA member states
and led by Principal Investigators from France and Italy,
telescope reflectors provided through a collaboration between ESA and a scientific consortium led and funded
by Denmark, and additional contributions from NASA
(USA). Some of the results in this paper have been derived using the \texttt{HEALPix} ~\cite{gorski/etal:2005} and \texttt{Astropy} ~\cite{2022ApJ...935..167A} packages.
\end{acknowledgments}
\bibliographystyle{apsrev4-2}
\bibliography{references}

\clearpage

\onecolumngrid
\begin{figure}[!t]
\centering
\includegraphics[width=\textwidth]{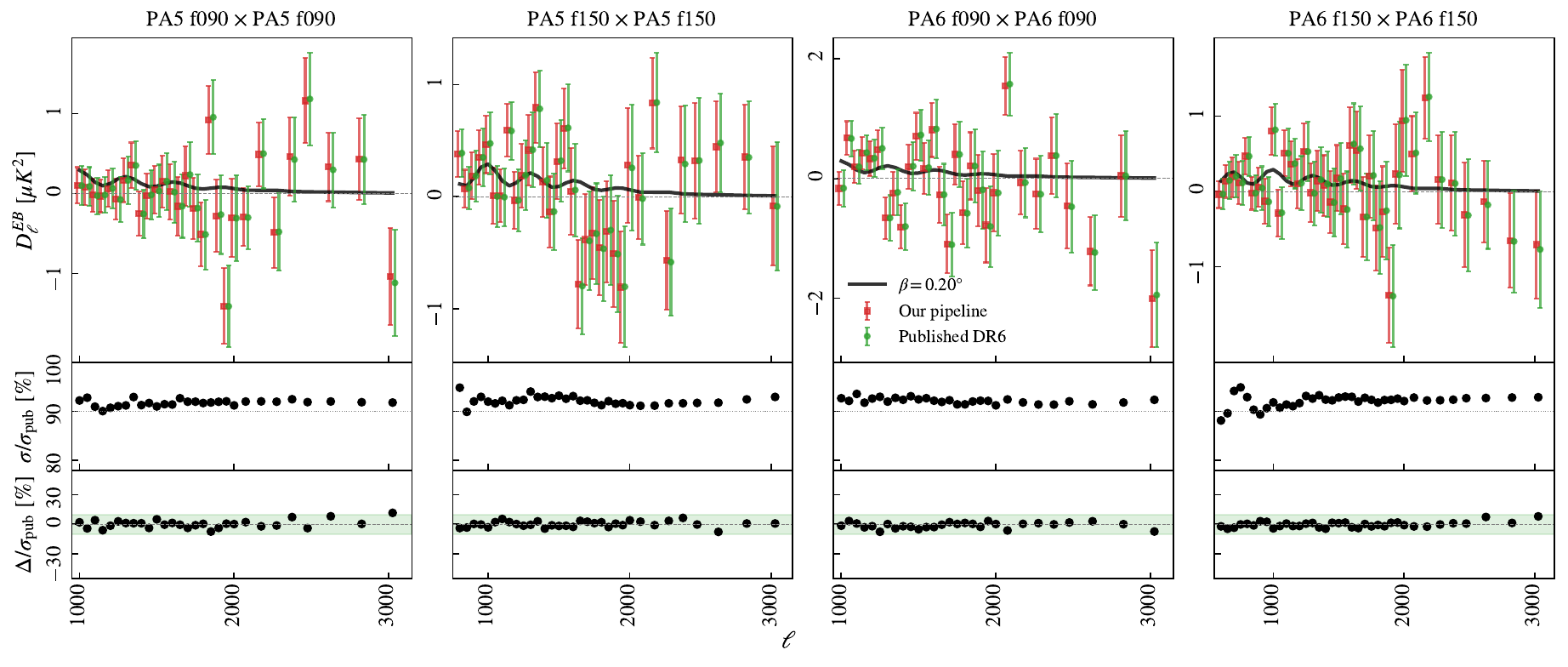}
\captionsetup{width=\textwidth,justification=raggedright,singlelinecheck=false}
\caption{\label{fig:spectra} Comparison of the published ACT DR6 and our pipeline spectra and covariance for a subset of the data. The released DR6 spectra are beam-convolved for this comparison. Top panel shows observed $EB$ spectra of our pipeline and the released data with uncertainties. Second row shows the ratio of the error bars between our pipeline and the released data, showing an average of $7\%$ underestimation of the uncertainty on $EB$. Third panel shows the difference in spectra normalized by the published error bar on $EB$ correlations. The green band shows $\pm 10\%$.
}
\end{figure}
\twocolumngrid

\section{Appendix A: Power spectrum and covariance estimation}

\subsection{Power spectrum}

To produce the power spectra and covariance elements that include ACT maps, we aim to closely follow the procedures of ACT outlined in Ref.~\cite{AtacamaCosmologyTelescope:2025blo} and in their codebase \footnote{\url{https://github.com/simonsobs/PSpipe}}$^{,}$\footnote{\url{https://github.com/simonsobs/pspipe_utils}}.

Following Section 3.3 and Appendix C of Ref.~\cite{AtacamaCosmologyTelescope:2025blo}, we correct for ground pickup by applying a k-space filter to remove modes with $|\ell_x| < 90$ and $|\ell_y| < 50$, where $x$ and $y$ are right ascension and declination, respectively. We start with the source-free map of the time-splits. To prevent the k-space filter from spreading bright point sources into the unmasked sky, we restore the point-source signal, defined as the difference between the original and source-free maps, only within the point-source mask. Any such spreading stays inside the region that will later be masked away.

With these k-space filtered maps, we use \texttt{pspy}\footnote{\url{https://github.com/simonsobs/pspy}} to compute the spectra with the full $\ell$-by-$\ell$ mode-coupling matrix and invert before binning. Since the time-splits have approximately uncorrelated noise~\cite{AtacamaCosmologyTelescope:2025vnj}, each cross-split spectrum is a signal-only estimator. Hence, we average over the $n_{\rm cross}$ different combinations of the time-split spectra to arrive at single-band spectra following Appendix C.4 in Ref.~\cite{AtacamaCosmologyTelescope:2025blo}. For ACT $\times$ \planck, we average only the ACT splits and not \planck\,splits. For example, to get $C^{E_{\rm PA5\,f090}B_{100A}}_{\ell}$, we average over $C^{E_{\rm PA5\,f090_i}B_{100A}}_{\ell}$ for split $i \in \{1,2,3,4\}$.

Then we apply the k-space filtering matrices $F^{-1}_b$ to the spectra to correct for the bias introduced by the filter. We are only interested in polarization, so we use the $3\times3$ elements of the matrices which describe $EE, BB, EB$ mixing. ACT also released these matrices for ACT $\times$ \planck\footnote{The released matrices cover ACT crossed with 100, 143, and 217\,GHz only. Because $F^{-1}_b$ is an ACT-side correction, it depends only on the ACT band, so we apply the same matrix to the remaining \planck\,channels.}. Lastly, we also subtract the residual temperature-to-polarization ($T\rightarrow P$) leakage using the released leakage coefficients $\gamma_{TE}$ and $\gamma_{TB}$. We perform these corrections on the ACT-leg of ACT $\times$ \planck\, spectra, too.

For the \planck\,$\times$\,\planck\, power spectra, we use the MASTER implemented pseudo-$C_\ell$ method \texttt{Namaster}\cite{Alonso:2018jzx, Hivon:2002}\footnote{\url{https://github.com/LSSTDESC/NaMaster}} as we find it to behave better at low $\ell$ compared to \texttt{pspy}. Using realistic \npipe\, simulations which included gain calibration, bandpass mismatch and beam systematics, Ref.~\cite{NPIPE:2022} found the bias on $\beta$ from known HFI systematics to be around $0.009^{\circ}$. As this is minor enough to neglect, we do not make corrections to the \planck\,spectra.

\subsection{Covariance}

The covariance matrix includes the correlations between all the ACT $\times$ ACT (AA), ACT $\times$ \planck\, (AP) and \planck\,$\times$ \planck\, (PP) spectra. For covariance in the PP-PP block (schematically Cov(PP, PP)), we have nearly full-sky coverage, and we compute these elements using an approximate covariance following previous \planck\, analyses \cite{minami/komatsu:2020b, NPIPE:2022, Eskilt:2022cff, Eskilt:2022wav}
\begin{equation}
    \label{eq:single-multipole-cov}
    \text{Cov}(C^{XY}_\ell, C^{ZW}_\ell) = \frac{C^{XZ,\textrm{o}}_\ell C^{YW,\textrm{o}}_\ell +  C^{XW,\textrm{o}}_\ell C^{YZ,\textrm{o}}_\ell}{(2\ell+1)f_{\textrm{sky}}}\,,
\end{equation}
where $f_{\rm sky}$ is the sky fraction of the mask.

Following previous work \cite{Eskilt:2022cff}, we drop every contraction on the right-hand side of Eq.\eqref{eq:single-multipole-cov} that contains a factor of $C_\ell^{EB,\, \rm o}$ due to statistical fluctuations. This reduces $\text{Cov}(C^{E_iB_j}_\ell, C^{E_pB_q}_\ell)$ to its $C_\ell^{E_iE_p} C_\ell^{B_jB_q}$ contribution only.

ACT maps have a smaller sky coverage than \planck, and so, for covariance elements that include at least one ACT mask (like Cov(AP, PP) or Cov(AA, PP)), we start with the analytic MASTER covariance \cite{Atkins:2024jlo, AtacamaCosmologyTelescope:2025blo} as our baseline. Instead of reducing the mask into a single $f_{\rm sky}$ it propagates the mode-coupling kernel $\Xi_{\ell\ell'}$ of the non-trivial ACT masks into the covariance elements. This also takes into account the very different sky coverages in the masks used for the spectra in each leg of Cov(AA, PP) and Cov(AP, PP). The covariance is first calculated for all per-split cross-spectra, before averaging over the $n_{\rm cross}$ time-split combinations for each spectrum-leg.

Although the analytic covariance matrix takes into account Gaussian noise and cosmic variance, it does assume homogeneous survey depth. ACT has complicated noise properties, so the team added simulation-based corrections, which we will not do.

However, we analytically correct the covariance from the ground pickup in two steps: first we sandwich the binned covariance elements with the $F^{-1}_b$ matrices. Secondly, Ref.~\cite{Atkins:2024jlo} pointed out that mode-count corrections needed to be applied to the covariance as approximately $\sqrt{F_b^{EE}}$ modes survive the filter. We, therefore, inflate the covariance by $(F_{b,i}^{EE}F_{b,j}^{EE})^{-1/4}$ for the covariance element of spectra $i$ and $j$. $F_b$ is simply the identity matrix for PP spectra in Cov(AA, PP) and Cov(AP, PP) elements as there is no ground pickup for \planck.

Besides noise and cosmic variance, we only correct for $T\rightarrow P$ uncertainty as this is the only additional error that contributes more than 1\% of the total error for polarization-only spectra (see Fig. 4 in Ref.~\cite{AtacamaCosmologyTelescope:2025blo}). We do this through adding an analytic covariance contribution from the uncertainty in the leakage coefficients $\gamma_{TE}$ and $\gamma_{TB}$.

After applying the corrections above, we validate our pipeline by comparing with the released DR6 $EB$ spectra and covariance for AA in Fig.~\ref{fig:spectra}. We find no bias in the spectra (bottom panels show the residual divided by the published error bars), while our error bars are understated by $\approx 7\%$. This aligns with the findings of Ref.~\cite{Atkins:2024jlo} which found that the analytical covariance underestimates the uncertainty. We find similar consistencies in $EE$ and $BB$.

As the \planck\, maps have simpler noise properties and no expected correlations with the noise of ACT, the covariance elements involving at least one Planck map are less affected by the approximations that drive the 7\% underestimate in the ACT-only block. We, therefore, expect the underestimation to be smaller for Cov(AP, AP), Cov(AP, AA), Cov(AP, PP), and Cov(AA, PP) than for Cov(AA, AA).

\end{document}